\documentclass[preprint2]{aastex}
\pdfoutput=1
\usepackage{graphicx}
\usepackage[bookmarks=true,breaklinks=true]{hyperref}

\begin{document}
\newcommand{\degree}{^{\circ}}
\newcommand{\text}[1]{\textrm{#1}}
\title{X-ray Variability and Evidence for Pulsations from the Unique Radio Pulsar/X-ray Binary Transition Object FIRST J102347.6+003841}
\shorttitle{X-ray Observations of J1023}
\shortauthors{Archibald et al.}
\author{Anne M. Archibald\altaffilmark{1}}
\email{aarchiba@physics.mcgill.ca}
\author{Victoria M. Kaspi\altaffilmark{1}}
\author{Slavko Bogdanov\altaffilmark{1}}
\author{Jason W. T. Hessels\altaffilmark{2,3}}
\author{Ingrid H. Stairs\altaffilmark{4}}
\author{Scott M. Ransom\altaffilmark{5}}
\and
\author{Maura A. McLaughlin\altaffilmark{6}}
\altaffiltext{1}{Department of Physics, McGill University, 3600 University St,  Montreal, QC H3A 2T8, Canada}
\altaffiltext{2}{Netherlands Institute for Radio Astronomy (ASTRON), Postbus 2, 7990 AA Dwingeloo, The Netherlands}
\altaffiltext{3}{Astronomical Institute ``Anton Pannekoek,'' University of Amsterdam, 1098 SJ Amsterdam, The Netherlands}
\altaffiltext{4}{Department of Physics and Astronomy, University of British Columbia, 6224 Agricultural Road, Vancouver, BC V6T 1Z1}
\altaffiltext{5}{National Radio Astronomy Observatory, 520 Edgemont Rd., Charlottesville, VA 22901, USA}
\altaffiltext{6}{Department of Physics, West Virginia University, Morgantown,   WV 26505, USA}

\begin{abstract}
We report on observations of the unusual neutron-star binary system \object[FIRST J102347.6+003841]{FIRST~J102347.6+003841} carried out using the \emph{XMM-Newton} satellite. This system consists of a radio millisecond pulsar in an $0.198$-day orbit with a $\sim 0.2\;M_\Sun$ Roche-lobe-filling companion, and appears to have had an accretion disk in 2001. We observe a hard power-law spectrum ($\Gamma = 1.26(4)$) with a possible thermal component, and orbital variability in X-ray flux and possibly hardness of the X-rays. We also detect probable pulsations at the pulsar period (single-trial significance $\sim 4.5\sigma$ from an $11(2)\%$ modulation), which would make this the first system in which both orbital and rotational X-ray pulsations are detected. We interpret the emission as a combination of X-rays from the pulsar itself and from a shock where material overflowing the companion meets the pulsar wind. The similarity of this X-ray emission to that seen from other millisecond pulsar binary systems, in particular \object[PSR J0024-7204w]{47~Tuc~W} (PSR~J0024$-$7204W) and \object[PSR J1740-5340]{PSR~J1740$-$5340}, suggests that they may also undergo disk episodes similar to that seen in J1023 in 2001.
\end{abstract}

\maketitle

\section{Introduction}

Radio millisecond pulsars are thought to have been spun up to their current rapid rotation by a long period of accretion in a Low-Mass X-ray Binary (LMXB). This general idea is well supported by many lines of evidence, but the transition between LMXB and radio millisecond pulsar remains mysterious. In particular, models of binary evolution often predict that mass transfer should continue indefinitely \citep{delo08}, but ionized material entering a neutron star's magnetosphere is expected to ``short out'' magnetospheric emission \citep{scc+94}. How then can we explain the observed radio millisecond pulsars?

An important system for studying this question was recently discovered: FIRST~J102347.6+003841 (hereafter J1023) is a 0.198-day binary containing a 1.69-ms radio millisecond pulsar, PSR~J1023+0038. This system is unique in that it shows evidence for having had an accretion disk as recently as 2001, but is now active as a radio pulsar \citep{asr+09,wat+09}. This suggests that we are observing the end of an LMXB phase and the birth of a radio millisecond pulsar, during which the system undergoes transient LMXB phases. This is consistent with a scenario put forward by \citet{bpd+01} in which LMXBs, late in their evolution, may have episodes of mass transfer but spend most of their time in ``radio ejection'' phases, during which the wind from an active radio MSP sweeps material overflowing the companion out of the system.

In its current quiescent state, J1023 has been extensively observed with optical and radio telescopes. The millisecond pulsar's companion appears as a $V\sim 17$ magnitude star with a G-like absorption-line spectrum. The brightness and temperature show a $0.198$-day modulation consistent with heating due to an irradiation luminosity of roughly $2 L_\sun$ from the primary \citep{ta05}. The combination of pulsar timing and optical radial velocity measurements shows that the mass ratio is $7.1(1)$, while assuming the pulsar mass is $\sim 1.4 M_\Sun$ implies an inclination angle of $\sim 46\degree$, a companion mass of $\sim 0.2\;M_\Sun$ and, based on Roche lobe size and optical apparent magnitude, a distance of $\sim 1.3\;\text{kpc}$ \citep{asr+09}. At radio wavelengths, the system shows variable, frequency-dependent eclipses near the orbital phase at which the companion is closest to our line of sight to the pulsar, though the companion's Roche lobe never actually intersects this line of sight. Irregular short eclipses are also observed at all phases \citep{asr+09}. Such variable radio eclipses have been observed in other millisecond pulsar systems, both in the so-called ``black widow'' systems in which the companion is thought to be a very low-mass degenerate star being ablated by the pulsar wind \citep{fst88,sbl+96}, and in several systems more similar to J1023, with companions that appear to be unevolved and Roche-lobe-filling \citep{fck+03,dlm+01}. Comparison of J1023 with these systems, as well as with LMXBs in quiescence, could clarify how accretion comes to an end and radio pulsars become active.

Prior to the discovery that J1023 contained a millisecond pulsar, and unfortunately after it had returned to quiescence, \citet{hsc+06} observed J1023 with \emph{XMM-Newton}. They concluded that it had a hard power-law spectrum, variability not necessarily correlated with the orbital phase, and a $0.5$--$10\;\text{keV}$ X-ray luminosity of $2.5\times 10^{32} \;\text{erg}\;\text{s}^{-1}$ ($0.065 L_\sun$) at a distance, based on estimates available at the time, of $2\;\text{kpc}$. This supported the suggestion of \citet{ta05} that J1023 had a NS primary rather than a white dwarf as had previously been thought. 

Not only does the detection of J1023 as a radio millisecond pulsar confirm that it is a NS binary, radio timing provides an ephemeris predicting the rotation of the pulsar. For this reason, we observed J1023 for a longer time and with high time resolution using \emph{XMM-Newton}. Our observations confirm the spectral observations of \citet{hsc+06}, show evidence for orbital variability, and show for the first time probable X-ray pulsations at the pulsar period.

In section 2 we summarize the observations we used and our initial data reduction. In section 3 we describe and give results from our spectral analysis and our search for orbital and rotational modulation in the X-ray data. In section 4 we discuss the implications for the nature of J1023's X-ray emission and the relation of J1023 to similar systems, and in section 5 we summarize and suggest further avenues of research.

\section{Observations}
The European Photon Imaging Camera, EPIC, aboard the \emph{XMM-Newton} satellite, has two MOS cameras \citep{taa+01} and one PN \citep{sbd+01}. All three cameras may be operated in an imaging mode, but the PN camera can also be run in a fast timing mode, in which the photons from one CCD are accumulated along one dimension, losing two-dimensional imaging, but achieving a time resolution of $29\;\mu\text{s}$.
Calibration uncertainties limit the absolute time accuracy of \emph{XMM-Newton} to approximately $100\;\mu \text{s}$ \citep{g+10}. \emph{XMM-Newton} also has a coaxial optical telescope, the Optical Monitor, and two reflection grating spectrometers, however J1023 is too faint to permit interesting studies with the latter. 

We obtained an \emph{XMM-Newton} observation (OBSID 0560180801) of J1023 of $33\;\text{ks}$, corresponding to $1.94$ binary orbits, on 2008 Nov 26. We operated the PN camera in fast timing mode and the MOS cameras in ``prime full window'' (imaging) mode, and we used the thin filter on all the EPIC cameras. The data were free of soft proton flares, so we were able to use the entire exposure time. We also retrieved the data set (OBSID 0203050201) used by \cite{hsc+06} from the \emph{XMM-Newton} archive; it consists of a single $16\;\text{ks}$ exposure ($0.92$ binary orbits) obtained on 2004 May 12, with all EPIC instruments using ``prime full window'' mode and the thin filter. This data set is also free of soft proton flares, but since all three EPIC instruments were operated in imaging mode, in this case high-resolution timing data are not available. 

The Optical Monitor was operated in both observations with the $B$ filter and a mode with $10$-s time resolution. Pipeline processing showed optical modulation consistent with that reported by \citet{ta05}, so we did no further analysis of the optical data.

We analyzed the X-ray data with the \emph{XMM-Newton} Science Analysis Software\footnote{The \textit{XMM-Newton} SAS is developed and maintained by the Science Operations Centre at the European Space Astronomy Centre and the Survey Science Centre at the University of Leicester.} version 9.0.0\footnote{\url{http://xmm.esac.esa.int/sas/}} and \texttt{xspec} version 12.5.0ac\footnote{\url{http://heasarc.nasa.gov/docs/xanadu/xspec/}}. 

We reprocessed the MOS and PN data with the \texttt{emchain} and \texttt{epchain} pipelines, respectively, then extracted photons meeting the recommended pattern, pulse invariant, and flag criteria. In particular, this restricted MOS photons to the $0.2$--$12\;\text{keV}$ energy range, PN imaging photons to the $0.13$--$15\;\text{keV}$ energy range, and PN timing photons to the $0.25$--$15\;\text{keV}$ energy range. For all analyses except the spectral fitting, we restricted the imaging photons from either type of camera to $0.2$--$10\;\text{keV}$ for consistency, since this involved discarding only a handful of photons. 

The default pipeline processing showed no evidence for extended emission from J1023, so we used the SAS tool \texttt{eregionanalyse} to select extraction regions to optimize the ratio of source photons to the square root of source plus background photons. These regions were circles of radius $35\arcsec$ (enclosing about $90\%$  and $85\%$ of the $1.5\;\text{keV}$ flux on the MOS and PN cameras, respectively) in imaging observations. For the timing-mode PN observation we adjusted the extraction region by hand to optimize detection significance (see below), selecting a band $12\arcsec$ wide (enclosing about $60\%$ of the $1.5\;\text{keV}$ flux). We selected background regions from nearby large circles in the imaging-mode observations, and from a band $50\arcsec$ wide, to one side of the source, in the timing-mode PN data. All together, this yielded approximately 5700 background-subtracted source photons in imaging mode and approximately 2400 background-subtracted source photons in timing mode. 

We computed photon arrival times using the SAS barycentering tool, \texttt{barycen}, the precise optical position $10^{\textrm{h}} 23^{\textrm{m}} 47.67^{\textrm{s}}$ $+0\degr 38\arcmin 41.2\arcsec$ (J2000) given in the NOMAD catalog \citep{zml+04}, and the DE405 solar system ephemeris. To produce pulsar rotational phases during the 2008 observation, we used the tool \texttt{tempo}\footnote{\url{http://www.atnf.csiro.au/research/pulsar/tempo/}} and the ephemeris given in \citet{asr+09}, which is based on radio pulsar timing data bracketing the X-ray observation epoch.  The extremely sensitive phase-coherent timing procedure used by \citeauthor{asr+09} revealed minuscule variations in the orbital period of J1023. These are modelled in the published ephemeris by expressing the orbital period as a quadratic polynomial obtained by a fit to several months of 2008 timing data. Extrapolating this polynomial back to the 2004 \emph{XMM-Newton} observation results in an unreasonable and poorly constrained orbital period at the time of $\sim -6$ days, a clear indication that the orbital period variations are not actually given by such a quadratic polynomial. While it is necessary to model the orbital period derivatives to obtain good quality \emph{pulsar} phase predictions, it is possible to obtain adequate \emph{orbital} phase predictions by using a simple model with constant orbital period of $0.19809620$~days and (pulsar) ascending node of MJD~$54801.970652993$, based on the model from \citet{asr+09}.  This model gives adequate orbital phase predictions back to 2004, matching the orbital phases observed in \citet{ta05} to within $1\%$ of a period as well as matching the ephemeris given in \citet{asr+09} exactly on 2008~Dec~1. We therefore used this model to evaluate the orbital phase of each barycentered photon in both our observations. 

\section{Results}

\subsection{Spectral fitting}
\begin{figure}
\plotone{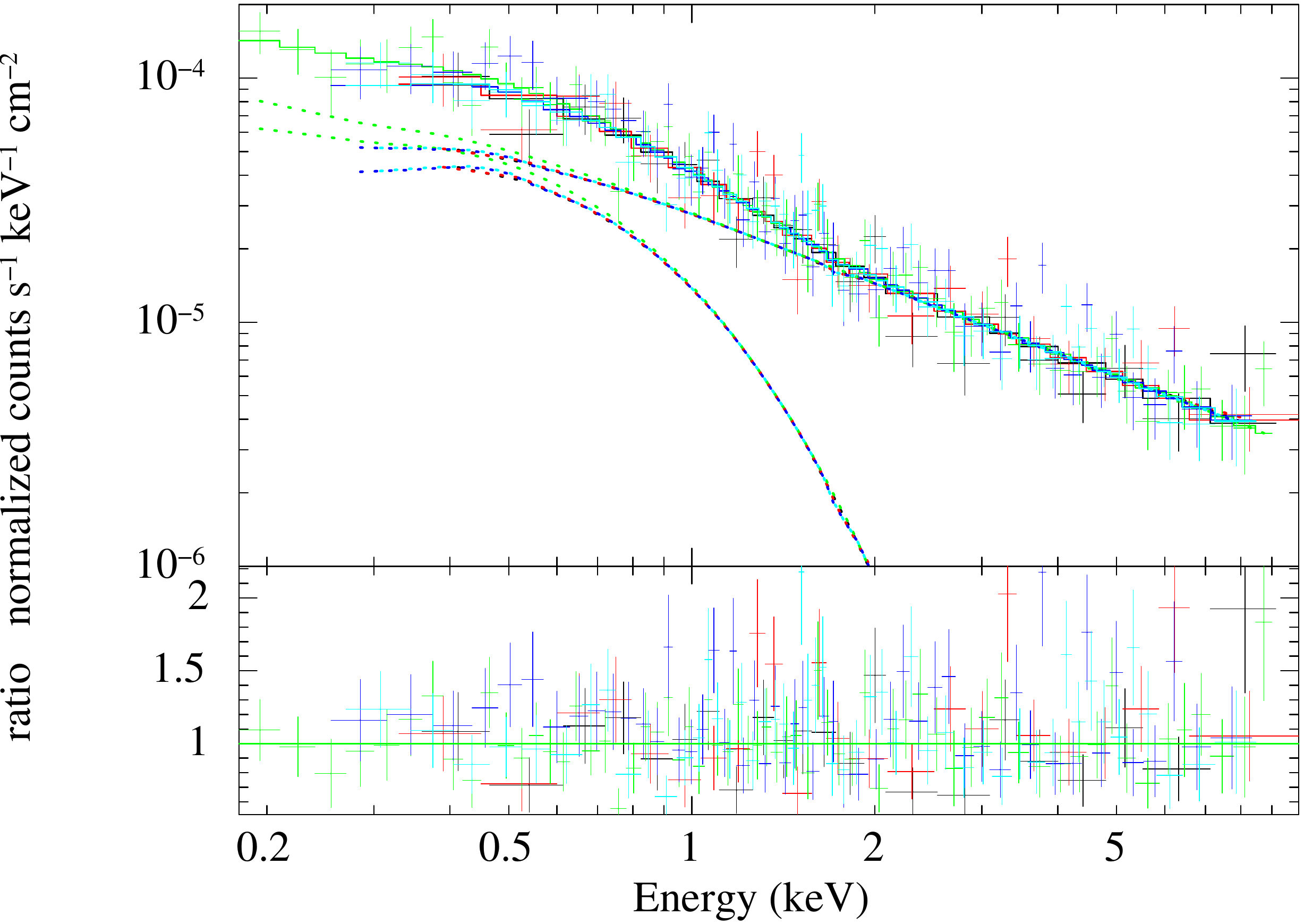}
\caption{\label{fig:spectrum}Absorbed neutron-star atmosphere plus power-law spectral fit to all EPIC imaging-mode photons from both observations. Each data set is plotted in a different color: black is the 2004 MOS1, red the 2004 MOS2, green the 2004 PN, blue the 2008 MOS1, and cyan the 2008 MOS2 data set. The vertical axis is normalized by the effective area, and the expected responses from the two additive components are overplotted separately (the harder component is the power law). The lower panel is the ratio of model predicted flux to observed flux in each group of detector channels.}
\end{figure}

\begin{deluxetable}{lcc}
\tablecaption{\label{table:fits}Spectral fits to all EPIC image data}
\tablewidth{0pt}
\tablecolumns{3}
\tablehead{
\colhead{} & \colhead{power-law} & \colhead{neutron star atmosphere } \\
\colhead{} & \colhead{} & \colhead{plus power law} }
\startdata
$N_{\textrm{H}}$ ($\text{cm}^{-2}$) 
& $<5\times 10^{19}$ & $<1\times 10^{20}$ \\
photon index 
& $1.26(4)$ & $0.99(11)$  \\
$kT$ ($\text{keV}$) / photon index ($\Gamma$) 
& \nodata & $0.12(2)$ \\
Thermal emission radius (km)\tablenotemark{a}
& \nodata & $0.7_{-0.1}^{+0.5}$ \\
$0.5$-$10\;\text{keV}$ unabsorbed flux ($\text{erg}\;\text{cm}^{-2}\;\text{s}^{-1}$) 
& $ 4.66(17)\times 10^{-13} $ & $ 4.9(3)\times 10^{-13} $ \\
$0.5$-$10\;\text{keV}$ luminosity\tablenotemark{a} ($\text{erg}\;\text{s}^{-1}$) 
& $9.4(4)\times 10^{31}$ & $9.9(5)\times 10^{31}$ \\
Thermal fraction 
& \nodata & $0.06(2)$ \\
$\chi^2$/degrees of freedom 
& 230/214 & 208/212 \\
Null hypothesis probability 
& 0.21 & 0.56 \\
\enddata
\tablenotetext{a}{Assuming a distance of $1.3\;\text{kpc}$.}
\end{deluxetable}

We used \texttt{xspec} to fit each of several models to all EPIC imaging data sets from both epochs simultaneously. To verify that the data sets were compatible, we also fit different absorbed power laws to the two observations, obtaining compatible spectral indices and normalizations. For the MOS cameras we restricted photons to the recommended energy range, namely $0.2$--$10\;\text{keV}$, and for the PN camera we used the recommended lower limit of $0.13\;\text{keV}$ but used the upper limit of $10\;\text{keV}$ as there were not enough photons above this to provide any constraint. We grouped the photons to obtain at least $20$ counts per spectral bin so as to have approximately Gaussian statistics.

We obtained an adequate fit to all imaging data sets with a power-law model including photoelectric absorption. We did not obtain adequate fits with an absorbed single black-body ($\chi^2=1203.7$ with $213$ degrees of freedom, for a null hypothesis probability of $\sim 10^{-138}$) or neutron-star atmosphere model ($\chi^2=1007.1$ with $213$ degrees of freedom, for a null hypothesis probability of $\sim 10^{-103}$).

The simple power-law model, with a photon index of $1.26(4)$, already has a null hypothesis probability of $0.21$, which is satisfactory from a statistical point of view. 
Although the statistics do not call for a thermal component, we may expect one on physical grounds (for example from the surface of the neutron star). We therefore also consider an absorbed neutron-star atmosphere plus power-law model to fit these data, as shown in Figure~\ref{fig:spectrum}. This model, specified in \texttt{xspec} as \texttt{phabs(nsa+pow)}, is based on \citet{zps96}. During fitting we froze the neutron star mass at $1.4 M_\sun$ and the radius at $10\;\text{km}$ for the purposes of redshift and light bending; the smaller effective thermal emission radius suggests that the emission comes from a small ``hot spot'' on the surface. In light of the fact that the estimated $B<10^8\;\text{G}$ \citep{asr+09}, we also selected a model in which the magnetic field has negligible effects on the atmosphere, i.e. $\lesssim 10^9\;\text{G}$. 

These two models are summarized in Table~\ref{table:fits}.  Uncertainties given are 90\% intervals returned by \texttt{xspec}'s \texttt{error} command; luminosities are obtained by using the \texttt{cflux} model component to estimate unabsorbed and thermal fluxes in the $0.5$--$10\;\text{keV}$ range and then using an assumed distance of $1.3\;\text{kpc}$ to determine a luminosity. ``Thermal fraction'' is the fraction of this luminosity due to the thermal component of the spectrum, if any. 

In both cases, fitting for photoelectric absorption in the model gave a very low upper bound on the neutral hydrogen column density, as noted in \citet{hsc+06}. Such a low value is to be expected since the entire Galactic column density in this direction is estimated\footnote{Using the HEASARC online calculator, \url{http://heasarc.nasa.gov/cgi-bin/Tools/w3nh/w3nh.pl}} to be only $1.9\times 10^{20}\;\text{cm}^{-2}$ \citep[based on][]{kbh+05}. 

Although both models are statistically adequate, the Akaike information criterion \citep{akai74} suggests one should prefer the model containing a thermal component. The $F$ test gives a null probability of $2\times 10^{-5}$ that the more complex model would produce such a large improvement in fit due to chance if the simple power-law model were correct. While there are concerns with using the $F$ test in such a situation (\citealt{pdc+02}, but see also \citealt{stew09}) it appears that the statistics do somewhat favor a model with a thermal component, although a purely non-thermal model cannot be excluded.

\subsection{Orbital variability}
\begin{figure}
\plotone{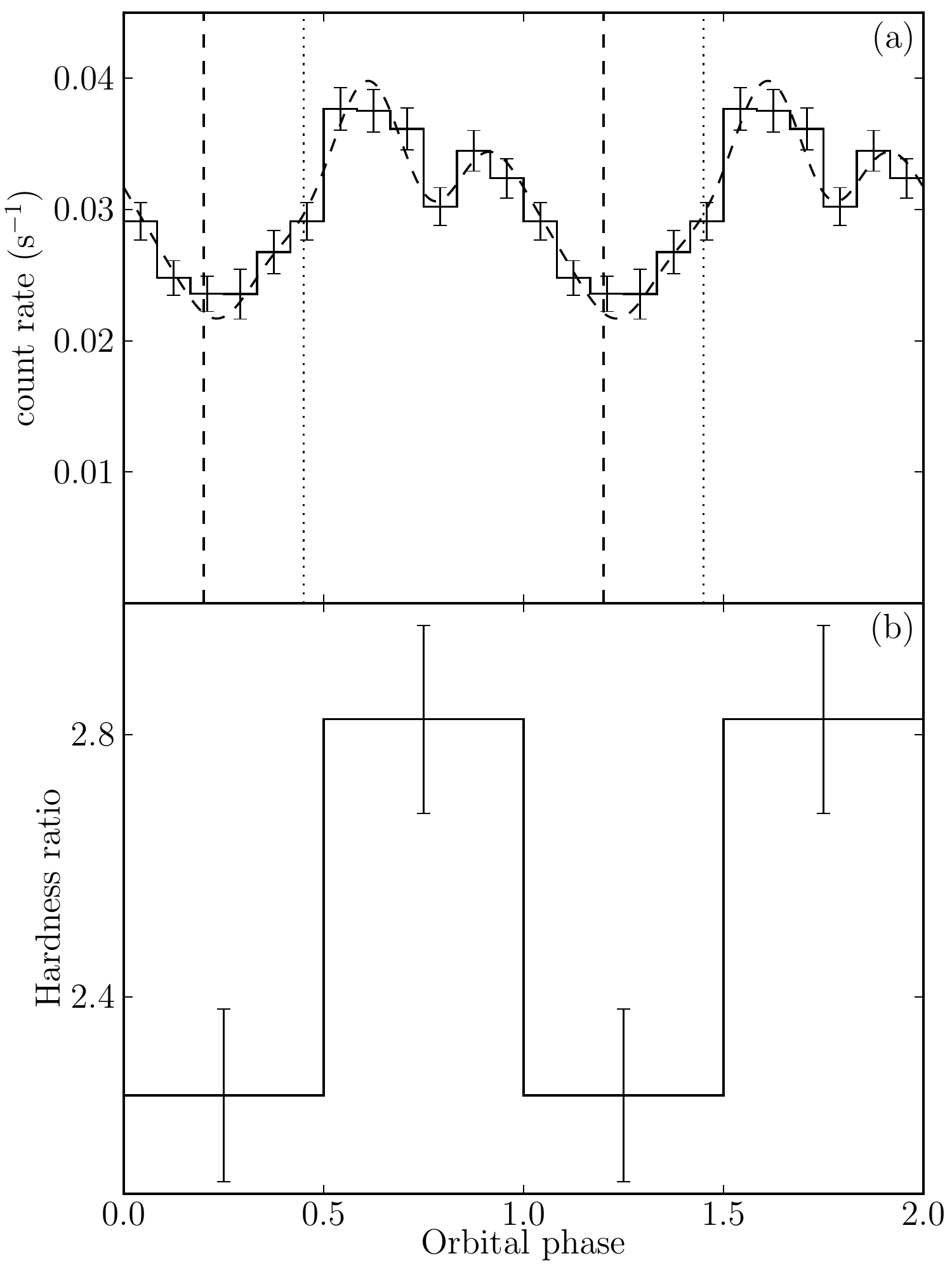}
\caption{\label{fig:orbitalplotfoldedmultiple}Panel (a): Average MOS-equivalent count rate in the $0.2$--$12\;\text{keV}$ energy range as a function of orbital phase. These data are a weighted average of all the imaging-mode data. Dashed and dotted vertical lines indicate the beginning and end of the $1.4\;\text{GHz}$ radio eclipse, respectively. The dashed curve is a best-fit combination of four sinusoids. Panel (b): number of MOS photons in the $1$--$12\;\text{keV}$ range divided by number of MOS photons in the $0.2$--$1\;\text{keV}$ range, also as a function of orbital phase. Both panels show two cycles for clarity. The companion's closest approach to our line of sight to the pulsar happens at orbital phase $0.25$; that the radio eclipse is off-center may be due to eclipsing material being far out of the orbital plane, as shown in Figure~\ref{fig:geometry}.}
\end{figure}
\begin{figure}
\begin{center}
\plotone{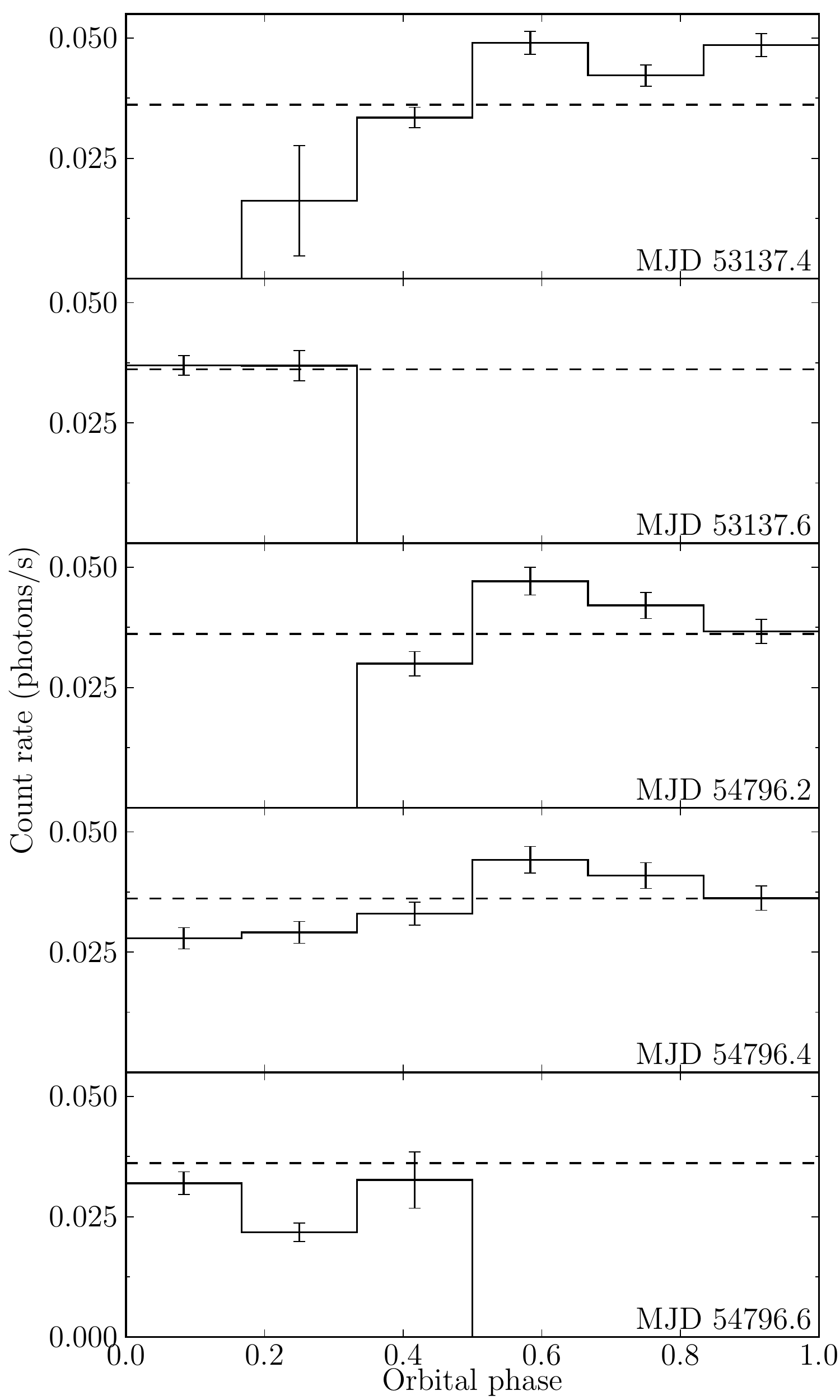}
\end{center}
\caption{\label{fig:orbitalplotall}Average count rate in the $0.2$--$12\;\text{keV}$ energy range as a function of orbital phase, for each orbit independently; the starting MJD for each orbit is indicated. The vertical axis is MOS-equivalent count rate (based on MOS1/2 and imaging-mode PN data, scaled to match the MOS1 count rate), and the dashed horizontal line indicates the average count rate over all orbital phases and data sets. Where no bar is plotted, no data are available.}
\end{figure}
To test for orbital variability, we selected $0.2$--$12\;\text{keV}$ source photons from the imaging-mode data sets in both our observations and those of \citet{hsc+06} and reduced them to the solar system barycenter. When combining data from multiple instruments, to take into account the different orbital coverages and sensitivities, we weighted bin values so that all average count rates match that observed in the MOS1 camera in the same observation. For the hardness ratio analysis, we omitted the PN data so that their different energy response and incomplete orbital coverage would not skew the results. 

A weighted histogram of these counts is shown in Fig.~\ref{fig:orbitalplotfoldedmultiple}, along with a best-fit sinusoid based on a finely binned profile. We selected four sinusoids as this appears to give a good representation of the curve and a time resolution similar to the histogram. A Kuiper test \citep{palt04} gives a probability of $1.3\times 10^{-19}$ ($9.0\sigma$) that photons drawn from a uniform distribution in orbital phase would be this non-uniform; the reduced $\chi^2$ for a constant fit to the histogram is $11.3$, with $11$ degrees of freedom. While the evidence for variability is strong, with only $\sim 3$ orbits covered by the two observations it is not certain that this variability is linked to orbital phase, although the fact that the minimum in the X-ray light curve occurs near orbital phase $0.25$, when the companion passes closest to our line of sight to the pulsar, suggests a link. To test this we plotted the photon arrival rates for each orbital period separately. Figure~\ref{fig:orbitalplotall} shows that the flux during each individual orbit appears to be lower during phase $0$--$0.5$ than during phase $0.5$--$1$, which suggests that the variability is indeed orbital. Note that \citet{hsc+06} detected variability, but having only limited orbital coverage, could not determine whether it was orbital. We also computed a hardness ratio (Fig.~\ref{fig:orbitalplotfoldedmultiple}b), dividing the number of photons harder than $1\;\text{keV}$ by the number of photons softer than $1\;\text{keV}$ and comparing the eclipse versus non-eclipse regions (for this purpose we defined the ``eclipse'' region to be phases $0$--$0.5$). The reduced $\chi^2$ for a fit of these values to a constant hardness ratio is $8.76$ for $1$ degree of freedom, and the probability of such a reduced $\chi^2$ arising if the hardness ratio were constant is $3.1\times 10^{-3}$ ($2.7\sigma$). Thus we see marginally significant softening in the eclipse region.

\subsection{Pulsations at the pulsar period}
\begin{figure}
\plotone{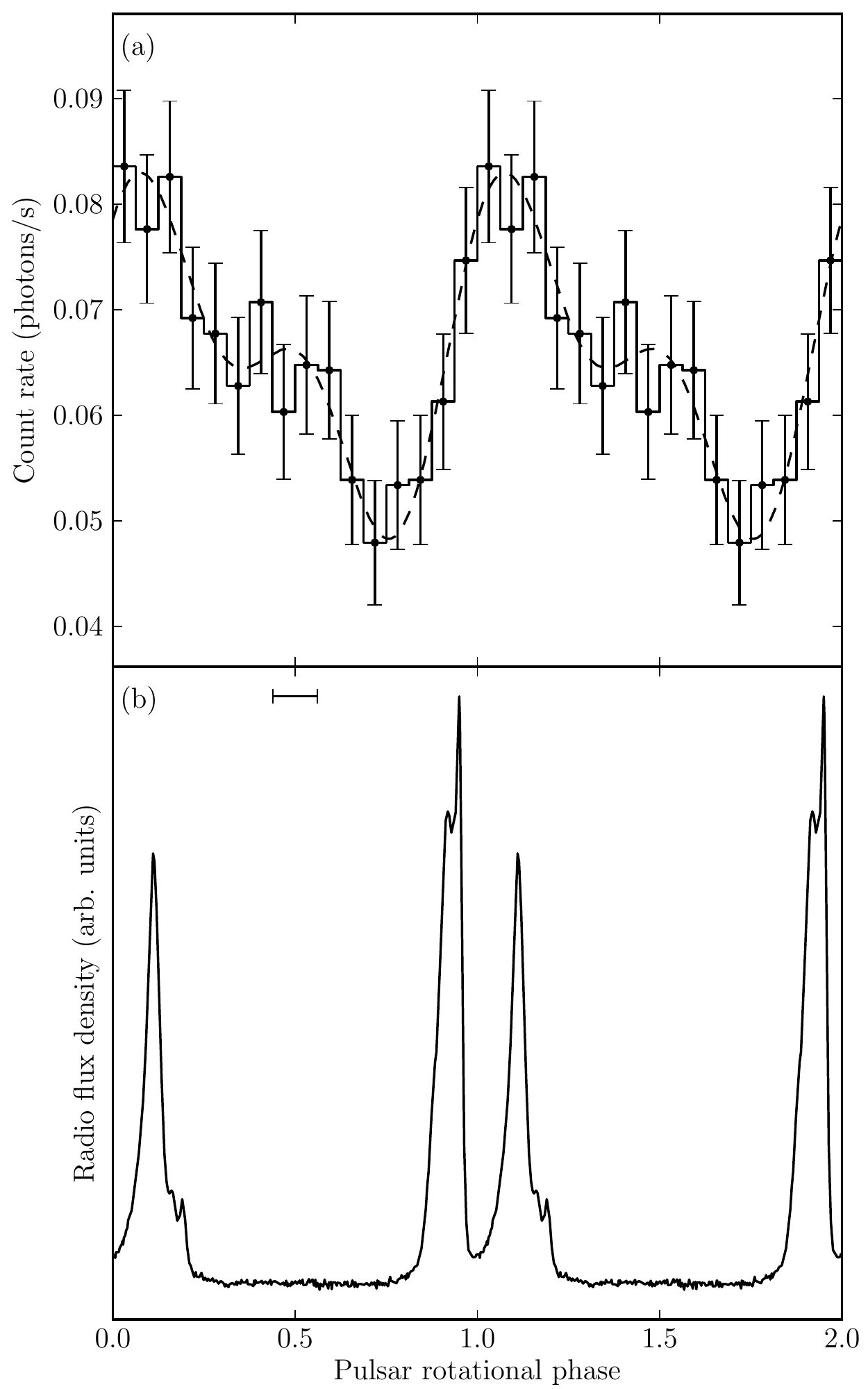}
\caption{\label{fig:pulsations}Panel (a) shows a background-subtracted light curve based on $0.25$--$2.5\;\text{keV}$ PN photons folded according to the radio ephemeris; the sinusoid (drawn with a dotted line) is a two-component sinusoid fit to the unbinned photon arrival times. Panel (b) shows a 1400 MHz profile obtained with Arecibo \citep{asr+09}. Both panels show two cycles for clarity.  The uncertainty in relative alignment, due primarily to the absolute timing uncertainty in the \emph{XMM-Newton} data, is indicated by the horizontal error bar. The  $20\;\mu\text{s}$ uncertainty relative to the radio data due to dispersion measure uncertainty is relatively unimportant.}
\end{figure}
To test for pulsations, we extracted source and background photons from the PN camera using the energy range $0.25$--$2.5\;\text{keV}$, selected to give the most significant detection. We barycentered the photon arrival times and used the program \texttt{tempo}\footnote{\url{http://www.atnf.csiro.au/research/pulsar/tempo/}} and the contemporaneous radio ephemeris given in \citet{asr+09} to assign each photon a rotational phase. We then tested these photons for uniform distribution in phase. The Kuiper test \citep{palt04} gave a (single-trial) null hypothesis probability of $3.7\times 10^{-6}$ ($4.5\sigma$) and the $H$ test \citep{jrs89} gave a (single-trial) null hypothesis probability of $2.4\times 10^{-6}$ ($4.6\sigma$), using an optimal number of sinusoids (two). We confirmed that no significant pulsations were detected with the background photons or with an incorrect ephemeris. We also verified that the period predicted by the ephemeris is very close to the period at which the significance peaks (holding all other ephemeris parameters fixed).

To estimate the degree to which the X-ray flux is modulated at the pulse period, one could simply take the lowest bin in the histogram as the background level and compute the fraction of photons above it; this yields a pulsed fraction of $0.27(9)$. However, this method is subject to large uncertainties, dependence on binning, and a large statistical upward bias due to the fact that we selected the bin in which signal plus noise is lowest, rather than that in which the (unknown) signal is lowest. To avoid these problems, we define a root-mean-squared pulsed flux by fitting a model $F(x)$ with two sinusoids:
\begin{displaymath}
f_{\text{RMS}} = \sqrt{\int_0^1 (F(x)-\bar F)^2 \text{d}x},
\end{displaymath}
where $\bar F$ is the model mean flux.  This is in some sense a degree of modulation; if the signal consists of a constant background plus a variable emission process that drops to zero, this is not directly measuring the fraction of emission due to the variable process. (There is a conversion factor that depends on the exact shape of the pulse profile.) However, this quantity can be computed with substantially less uncertainty and bias. Computationally, we estimate the root-mean-squared amplitude in the Fourier domain, based on the amplitudes of the two complex Fourier coefficients (since higher-order Fourier coefficients are dominated by noise); since noise always contributes a positive power to Fourier coefficients, to reduce the bias we subtract the expected contribution of noise from the squared amplitude of each coefficient before taking the square root. We then convert this pulsed flux value to a pulsed fraction by dividing by the total background-subtracted flux from J1023.  

For the energy range $0.25$--$2.5\;\text{keV}$ we estimate a root-mean-squared pulsed fraction of $0.11(2)$. In the two subbands $0.25$--$0.6\;\text{keV}$ and $0.6$--$2.5\;\text{keV}$, we find root mean squared pulsed fractions of $0.17(5)$ and $0.14(3)$, respectively, with profiles that are broadly similar and in phase. Above $2.5\;\text{keV}$ we find no evidence for pulsations, though a $3\sigma$ upper limit on the pulsed fraction is only $0.20$.

\section{Discussion}

To summarize our results for J1023, we observed an X-ray spectrum dominated by a hard power-law component, possibly with a small, soft, thermal contribution. The emission appears to be modulated at the $0.198$-day orbital period, with substantial dips as the companion passes near the line of sight; these dips are accompanied by a possible softening in the spectrum. We also found that the X-ray emission is very likely modulated at the $1.69$-ms rotational period of the radio pulsar. 

\subsection{Nature of the X-ray emission}

In a review of X-ray emission from millisecond pulsars, \citet{zavl07} describes three primary sources of X-ray emission: emission from an intrabinary shock, emission from the neutron star itself, which is some combination of thermal and magnetospheric emission, and emission from a pulsar wind nebula outside the binary system. As all these mechanisms may be operating in J1023, we next consider their contributions, if any, to the observed X-rays from this system.

\begin{figure}
\plotone{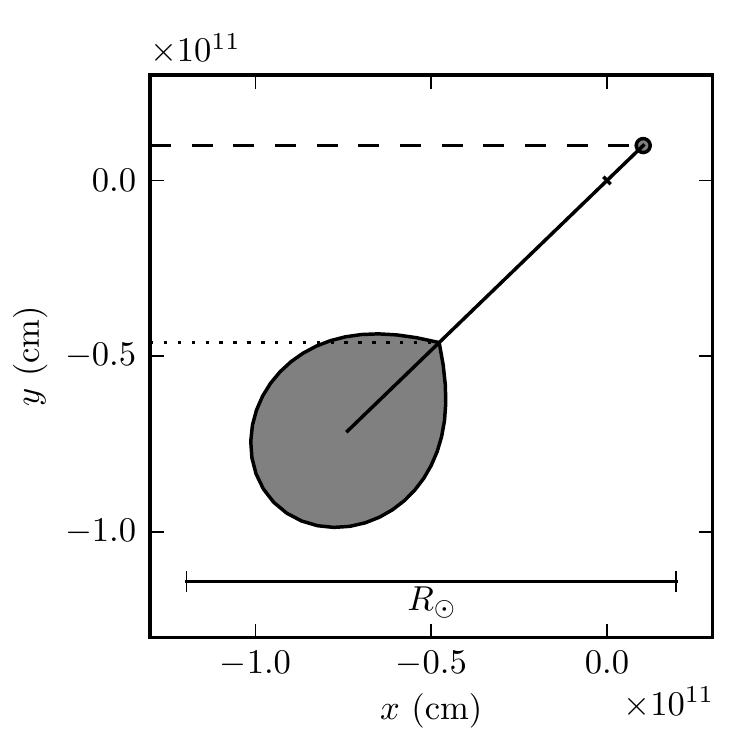}
\caption{\label{fig:geometry}J1023 system geometry at phase $0.25$ as seen from within the plane of the sky. The system orbit is in and out of the page, the dashed line is our line of sight to the pulsar, and the dotted line is our line of sight to L1, which is eclipsed by the Roche-lobe-filling companion. The pulsar is indicated by a not-to-scale circle, but the companion is drawn to scale. The tick mark on the line connecting pulsar and companion marks the center of mass of the system. This system geometry assumes a pulsar mass of $1.4\;M_{\Sun}$; a higher mass would mean a larger system seen more nearly face-on, but the range of plausible angles is constrained to $\sim 53\degree$--$34\degree$ by the narrow range of plausible pulsar masses $1.0$--$3.0M_\Sun$ \citep{asr+09}. Note that at this orbital phase the X-rays are near their minimum, and low radio frequencies from the pulsar are eclipsed, indicating the presence of ionized material well out of the orbital plane.}
\end{figure}
\subsubsection{Emission From an Intrabinary Shock}
J1023 likely has a strong pulsar wind, since the optical data suggest that the companion is being heated by a luminosity of $\sim 2 L_\Sun$ from the pulsar \citep{ta05}. This is much greater than the X-ray luminosity of the system ($\sim 0.03 L_\Sun$), but well below the upper limit ($\sim 80 L_\Sun$) on the spin-down luminosity of the pulsar \citep{asr+09}. If material is leaving the companion, either through Roche-lobe overflow or a stellar wind, we should expect an intrabinary shock, where this flow meets the pulsar wind. Such a shock could readily produce power-law X-ray emission \citep{at93}. If localized, it could easily account for the orbital modulation we observe in the X-ray emission. If the emission were due to Roche-lobe-overflowing material meeting the pulsar wind we might expect it to be localized at or near L1. Given the system geometry described in \citet{asr+09} (a $46\degree$ inclination angle, corresponding to a pulsar mass of $1.4 M_\Sun$, and a Roche-lobe-filling companion; see Figure~\ref{fig:geometry}), the L1 point itself is eclipsed by the companion for $0.32$ of the orbit, centered on orbital phase $0.25$. 
We would expect spectral softening during this period, since the relatively hard power-law emission is blocked but the, presumably softer, emission from the neutron star is not. A larger X-ray-emitting region, due either to material streaming away from the L1 point or to a stellar wind shock, would result in a broader, shallower, and potentially asymmetric dip in the X-ray light curve. 

The possibility of a wind from the companion is interesting, since it is not clear how Roche-lobe overflow would provide the ionized material causing the radio eclipses, which occur well above the orbital plane, roughly centered on the companion's closest approach to our line of sight. \citet{asr+09} reported dispersion measure (DM) changes of $\sim 0.15\;\text{pc}\;\text{cm}^{-3}$ above a cutoff frequency of $\sim 1\;\text{GHz}$ at radio eclipse ingress and egress. Assuming that this is the plasma frequency implies an electron density of $\sim 10^{10}\;\text{cm}^{-3}$; combined with the excess DM measurement, this implies a layer of thickness $\sim 4\times 10^{7}\;\text{cm}$, relatively thin compared to the orbital separation of $1.2\times 10^{11}\;\text{cm}$. This thin sheet of material could in principle arise as the shock front where the pulsar wind meets a wind from the companion, but it is difficult to explain the companion's wind being strong enough to support such a shock: given the line of sight geometry of the system, the material causing the radio eclipses must be about as far from the companion as it is from the pulsar. If we assume that the highly relativistic pulsar wind provides the $\sim 2 L_\Sun$ required by the companion heating models of \citet{ta05}, it seems difficult to explain how the companion could have a wind of comparable pressure anywhere along the line of sight. Thus the radio eclipses remain a mystery, and Roche-lobe overflow seems a more likely explanation for the origin of the intrabinary shock.

\subsubsection{Emission From the Neutron Star}
Some millisecond pulsars, binary or isolated, have X-ray emission modulated at the rotational period. This emission is some combination of magnetospheric emission, from high-energy particles in the magnetosphere, and thermal emission, from polar caps heated by bombardment by high-energy particles moving along the open field lines. Observationally, \citet{zavl07} draws the distinction that magnetospherically dominated emission has high pulsed fractions, narrow peaks, and power-law spectra, while polar-cap-dominated emission has lower pulsed fractions, broader peaks, and thermal spectra. \citeauthor{zavl07} also suggests that pulsars with spin-down luminosities $\gtrsim 10^{35}\;\text{erg}\;\text{s}^{-1}$ tend to be magnetospherically dominated while those with spin-down luminosities $\lesssim 10^{34}\;\text{erg}\;\text{s}^{-1}$ tend to be thermally dominated, but since few examples are known, this classification is tentative, and the spin-down luminosity from J1023 is only known to be $<3\times 10^{35}\;\text{erg}\;\text{s}^{-1}$ \citep{asr+09}. In any case, classification of the pulsations we observe in J1023 is difficult. 

In systems in which thermal emission from the neutron star provides all the detectable X-rays, detailed modelling indicates that the pulsed fraction should normally be $\lesssim 50\%$ (that is, pulsed emission must be accompanied by roughly equal or greater unpulsed emission) due to light bending and the large size of the polar caps, though for certain combinations of parameters higher pulsed fractions can occur \citep{bgr08}. If the pulsations are purely thermal, then our spectral upper limit on the thermal fraction of $0.06(2)$ is difficult to reconcile with the fact that we observe a fractional modulation of $0.11(2)$ at the pulsar period.  Thus it seems likely that at least some magnetospheric emission is present, as it has a nonthermal spectrum and can more readily have higher pulsed fractions. The poor signal-to-noise in our observations, due to the high background (from the system itself as well as instrumental) and the scarce photons, makes it impractical to determine the sharpness of the pulse profile or the hardness of the pulsations. In any case the pulsations, if real, are a clear sign of X-ray emission from the pulsar itself.

\subsubsection{Emission From a Pulsar Wind Nebula}
Nebular emission from pulsars arises when the particle wind driven by the pulsar's magnetospheric activity flows out of the pulsar's immediate neighborhood and meets the surrounding medium; for reviews see \citet{gs06} or \citet{krh06}. 
Pulsars like J1023 whose wind is confined by the ambient interstellar medium exhibit nebular emission that generally takes a cometary form, with an arc-like bow shock preceding the pulsar and a ``trail'' of ejected material streaming back along the pulsar's track. 

The angular size of bow shock emission depends on the pulsar's spin-down luminosity, its proper motion, and the local interstellar medium density. \citet{kp08} give a formula for predicting the ``stand-off'' angle of the X-ray bow shock from the pulsar:
\[
\theta = 5.4\arcsec\;n_{0.1}^{-1/2}\mu_{19}^{-1}{\dot E_{35}}^{1/2}D_{1.3}^{-2},
\]
where $n_{0.1}$ is the mean density of the interstellar medium divided by $0.1\; \text{cm}^{-3}$, $\mu_{19}$ is the proper motion divided by $19\;\text{mas}\;\text{yr}^{-1}$, $\dot E_{35}$ is the spin-down luminosity divided by $10^{35}\;\text{erg}\;\text{s}^{-1}$, and $D_{1.3}$ is the distance to J1023 divided by $1.3\;\text{kpc}$. We have supplied best-guess parameters for J1023 \citep{asr+09}, so that absent further information, we might expect a stand-off distance of $\sim 5\arcsec$ for the X-ray bow shock.  Thus although nebular emission is not resolved in our \emph{XMM-Newton} images, higher-resolution images with the \emph{Chandra X-ray Observatory}, or in the $H\alpha$ or radio bands, might yet resolve a bow shock. 

\citet{kp08} find that the ratio of neutron-star emission to nebular emission generally lies between $\sim 0.1$ and $10$, which suggests that nebular emission may produce some of the X-rays we observe.  Moreover, \citeauthor{kp08} find that nebular emission from bow shocks has a power-law spectrum with photon indices $1\lesssim \Gamma\lesssim 2$ \citep{kp08}, consistent with what we observe from J1023, though the latter is somewhat harder than most pulsar wind nebulae. The orbital variability, on the other hand, argues against that component of the emission arising from an extrabinary pulsar wind nebula.

\subsection{Comparison to similar systems}

In quiescence, most LMXBs have spectra that are dominated by thermal emission from the neutron star. This emission is typically consistent with a hydrogen atmosphere model with effective radius $\sim 10\;\text{km}$ \citep{bbr98}. Some LMXBs in quiescence also have a small power-law component of photon index $1\lesssim \Gamma \lesssim 2$ in their spectrum \citep{ccm+98}, while a few quiescent LMXBs have spectra completely dominated by this non-thermal component. For example, the quiescent LMXB \object[SAX J1808.4-3658]{SAX~J1808.4$-$3658} has a spectrum that is a fairly hard power law with no detectable thermal emission \citep{hjw+09}. J1023 resembles this latter category, although if a thermal component is present its effective radius appears to be substantially less than $10\;\text{km}$ and the power-law spectral index is fairly hard. On the other hand, J1023's radio and possible X-ray pulsations distinguish it from all known quiescent LMXBs.

SAX~J1808.4$-$3658 is of particular interest because it is the prototype and best-studied example of a class of LMXBs from which millisecond X-ray pulsations have been detected during active phases. Many authors have suggested that SAX~J1808.4$-$3658 should turn on as a radio pulsar during quiescence \citep[e.g.][]{wk98,cm98,cdc+04}, but no radio pulsations have been detected in spite of thorough searches \citep[e.g.][]{bbp+03,ibb+09}. The non-detection of radio pulsations could be because SAX J1808.4$-$3658 is shrouded by previously ejected ionized material, or it could simply be an issue of unfortunate beaming.  In any case, it is natural to compare SAX J1808.4$-$3658 to J1023. 

During its active phases, SAX J1808.4$-$3658 reaches a sustained luminosity of $\sim 4\times 10^{36}\;\text{erg}\;\text{s}^{-1}$ \citep{gc06}, in sharp contrast to the upper limit of $2\times 10^{34}\;\text{erg}\;\text{s}^{-1}$, available for J1023 during its disk phase \citep{asr+09}. In its quiescent state, SAX J1808.4$-$3658 shows X-ray emission dominated by a hard power law ($\Gamma = 1.74(11)$), with a small ($\lesssim 0.1$) fraction of thermal emission, and a total X-ray luminosity of $7.9(7)\times 10^{31}\;\text{erg}\;\text{s}^{-1}$ \citep{hjw+09}. This is somewhat similar to what we observe in J1023, which has a total luminosity of $9.9(5)\times 10^{31}\;\text{erg}\;\text{s}^{-1}$, although the spectral index of $1.26(4)$ is somewhat harder than that seen from SAX J1808.4$-$3658. If J1023 has a thermal component, its power-law is substantially harder, with a spectral index of $0.99(11)$.  The thermal component would have an effective radius of $0.7_{-0.1}^{+0.5}\;\text{km}$ and supply $0.06(2)$ of the total luminosity. While the spectral index seen in J1023 is harder than that seen from SAX J1808.4$-$3658, it is in line with the hard spectral indices seen from other X-ray-detected radio millisecond pulsars (for example 47~Tuc~W, though PSR~J1740-530 is softer; see below).  Since SAX~J1808.4$-$3658 is known to cool quickly, even thermal X-ray pulsations in quiescence would be evidence of magnetospheric activity heating its polar caps.  \citet{csg+02} searched for such X-ray pulsations from SAX J1808.4$-$3658 in quiescence, but were unable to conduct meaningful searches due to its faintness; SAX~J1808.4$-$3658 is roughly $3.5\;\text{kpc}$ away \citep{gc06} compared to the roughly $1.3\;\text{kpc}$ we assume for J1023 \citep{asr+09}.

The similarity of the X-ray properties of SAX J1808.4$-$3658 in quiescence to those of J1023 suggests that SAX J1808.4$-$3658 may indeed harbor a radio millisecond pulsar in quiescence; conversely, the same similarity suggests that J1023 may undergo episodes of active accretion.

Among radio pulsars, the two best-studied examples which closely resemble J1023 are 47~Tuc~W \citep{bgb05} and PSR~J1740$-$5430 \citep{dpm+01}. Both pulsars exhibit large, variable radio-frequency eclipses, and both are thought to have somewhat massive unevolved companions ($\gtrsim 0.13 M_\Sun$ and $0.19$--$0.8\;M_\Sun$, respectively). For comparison, J1023's companion is thought to be $\sim 0.2\;M_\Sun$, and shows a spectrum characteristic of a main-sequence star \citep{asr+09,ta05}. The radio eclipses are in all three cases evidence for material leaving the companion and presumably being expelled from the system, and in all three cases the companion appears to fill, and perhaps overflow, its Roche lobe \citep{bgb05,fpds01,ta05}. 

\citet{bgb05} studied 47 Tuc W and found an X-ray spectrum consisting of a dominant hard power-law ($\Gamma=1.14(35)$, contributing $\sim 75\%$ of the total flux) plus a thermal component. They also observed orbital variability, and in particular a substantial X-ray eclipse spanning phases\footnote{Note that these authors use a different convention for orbital phase, setting phase 0.5 to the optical minimum, while we set phase 0.25 to the optical minimum. All phases given here have been converted to our convention.} $0.15$--$0.45$ and centered slightly ahead of the optical minimum, at phase $0.2$. This is accompanied by substantial softening of the X-ray spectrum during eclipse. Given these observations, \citeauthor{bgb05} suggest that the power-law X-ray emission originates from a shocked region where material overflowing from the companion of 47 Tuc W meets the pulsar wind and is blown out of the system. 
\citet{bbh+10} studied \emph{Chandra X-ray Observatory} observations of PSR~J1740$-$5430. Although limited by the scarcity of photons, they found a spectrum consistent with a power law of index $\Gamma=1.73(8)$, or with a somewhat harder power-law plus a black body component. Their estimates of variability were limited by restricted orbital coverage as well as by a paucity of photons, but they found marginal evidence for orbital modulation, in particular a possible decrease in luminosity during the radio eclipses, possibly accompanied by a softening. 
Our observations of J1023 are more photon-starved than those of \citet{bgb05} of 47~Tuc~W, but we do see evidence for eclipses, and possibly for softening during eclipses. In the case of J1023, the possibility of overflowing gas is supported by the recent disk phase. 

The similarity of the observational properties of these three millisecond pulsars suggests that their companions are all overflowing their Roche lobes, so that mass flows through the L1 point, where it encounters the pulsar wind and is swept away. On the other hand, the similarity also suggests the possibility that like J1023, 47~Tuc~W or PSR~J1740$-$5430 may undergo episodes in which the mass transfer rate from the companion increases and an accretion disc forms. Such episodes would presumably be signalled by optical changes, possibly by extinction of the radio pulsations, and presumably by modest X-ray brightening. All together, these observations suggest that as a system reaches the end of its life as an LMXB, its accretion rate drops and the millisecond pulsar becomes active. After this point, the wind from the pulsar generally prevents mass leaving the companion from forming an accretion disk or entering the pulsar magnetosphere. Temporary increases in the mass accretion rate may allow the occasional suppression of the wind and formation of an accretion disk, as was observed in J1023 in 2001.

\section{Conclusions}

J1023 shows X-ray emission probably consisting primarily of emission from an intrabinary shock, plus a smaller amount of emission from the pulsar itself and possibly some unresolved nebular emission. The shock emission shows substantial variability that appears to be linked to orbital phase, as well as possible spectral softening during eclipses.  Similarities to 47~Tuc~W and PSR~J1740$-$5340 suggest that this may be due to gas overflowing from the companion meeting the pulsar wind in a shock close to the companion.  The emission from the pulsar itself, if real, is probably due to some combination of curvature radiation in the magnetosphere and thermal radiation from polar caps heated by high-energy particles streaming downward, though which effect dominates is unclear. Further observations to provide improved spectra should help clarify the origin of both pulsed and unpulsed emission, as would further high-time-resolution observations to allow better measurement of pulse profiles, pulsed hardness ratios, and orbital variations of pulsed fraction. X-ray observations with higher spatial resolution might also resolve extended nebular emission around J1023. Better understanding of this object promises to clarify the transition from LMXB to radio millisecond pulsar, and may show that other known systems are currently in this fascinating transition state.

\section*{Acknowledgements}
The authors thank Mallory S. E. Roberts for useful discussions.
This work was based on observations obtained with \textit{XMM-Newton}, an ESA science mission with instruments and contributions directly funded by ESA Member States and NASA. 
AMA is supported by a Schulich graduate fellowship.
VMK acknowledges support from NSERC, FQRNT, CIFAR, and holds a Canada
Research Chair and the Lorne Trottier Chair in Astrophysics and Cosmology.
SB is a CIFAR Junior Fellow. 
JWTH is an NWO Veni Fellow.  
IHS received support from an NSERC Discovery Grant.
MAM is supported by a WV EPSCOR Research Challenge Grant.  

\emph{Facility:} \emph{\facility{XMM-Newton}}

\bibliographystyle{apj}
\bibliography{journals,refs}{}

\begin{thebibliography}{42}
\expandafter\ifx\csname natexlab\endcsname\relax\def\natexlab#1{#1}\fi

\bibitem[{{Akaike}(1974)}]{akai74}
{Akaike}, H. 1974, IEEE Transactions on Automatic Control, 19, 716

\bibitem[{{Archibald} {et~al.}(2009){Archibald}, {Stairs}, {Ransom}, {Kaspi},
  {Kondratiev}, {Lorimer}, {McLaughlin}, {Boyles}, {Hessels}, {Lynch}, {van
  Leeuwen}, {Roberts}, {Jenet}, {Champion}, {Rosen}, {Barlow}, {Dunlap}, \&
  {Remillard}}]{asr+09}
{Archibald}, A.~M., {Stairs}, I.~H., {Ransom}, S.~M., {Kaspi}, V.~M.,
  {Kondratiev}, V.~I., {Lorimer}, D.~R., {McLaughlin}, M.~A., {Boyles}, J.,
  {Hessels}, J.~W.~T., {Lynch}, R., {van Leeuwen}, J., {Roberts}, M.~S.~E.,
  {Jenet}, F., {Champion}, D.~J., {Rosen}, R., {Barlow}, B.~N., {Dunlap},
  B.~H., \& {Remillard}, R.~A. 2009, Science, 324, 1411

\bibitem[{{Arons} \& {Tavani}(1993)}]{at93}
{Arons}, J. \& {Tavani}, M. 1993, \apj, 403, 249

\bibitem[{{Bogdanov} {et~al.}(2008){Bogdanov}, {Grindlay}, \&
  {Rybicki}}]{bgr08}
{Bogdanov}, S., {Grindlay}, J.~E., \& {Rybicki}, G.~B. 2008, ApJ, 689, 407

\bibitem[{{Bogdanov} {et~al.}(2005){Bogdanov}, {Grindlay}, \& {van den
  Berg}}]{bgb05}
{Bogdanov}, S., {Grindlay}, J.~E., \& {van den Berg}, M. 2005, ApJ, 630, 1029

\bibitem[{{Bogdanov} {et~al.}(2010){Bogdanov}, {van den Berg}, {Heinke},
  {Cohn}, {Lugger}, \& {Grindlay}}]{bbh+10}
{Bogdanov}, S., {van den Berg}, M., {Heinke}, C.~O., {Cohn}, H.~N., {Lugger},
  P.~M., \& {Grindlay}, J.~E. 2010, \apj, 709, 241

\bibitem[{{Brown} {et~al.}(1998){Brown}, {Bildsten}, \& {Rutledge}}]{bbr98}
{Brown}, E.~F., {Bildsten}, L., \& {Rutledge}, R.~E. 1998, \apjl, 504, L95+

\bibitem[{{Burderi} {et~al.}(2001){Burderi}, {Possenti}, {D'Antona}, {Di
  Salvo}, {Burgay}, {Stella}, {Menna}, {Iaria}, {Campana}, \&
  {d'Amico}}]{bpd+01}
{Burderi}, L., {Possenti}, A., {D'Antona}, F., {Di Salvo}, T., {Burgay}, M.,
  {Stella}, L., {Menna}, M.~T., {Iaria}, R., {Campana}, S., \& {d'Amico}, N.
  2001, \apjl, 560, L71

\bibitem[{{Burgay} {et~al.}(2003){Burgay}, {Burderi}, {Possenti}, {D'Amico},
  {Manchester}, {Lyne}, {Camilo}, \& {Campana}}]{bbp+03}
{Burgay}, M., {Burderi}, L., {Possenti}, A., {D'Amico}, N., {Manchester},
  R.~N., {Lyne}, A.~G., {Camilo}, F., \& {Campana}, S. 2003, \apj, 589, 902

\bibitem[{{Campana} {et~al.}(1998){Campana}, {Colpi}, {Mereghetti}, {Stella},
  \& {Tavani}}]{ccm+98}
{Campana}, S., {Colpi}, M., {Mereghetti}, S., {Stella}, L., \& {Tavani}, M.
  1998, Astron.\ Astropys.\ Rev., 8, 279

\bibitem[{{Campana} {et~al.}(2004){Campana}, {D'Avanzo}, {Casares}, {Covino},
  {Israel}, {Marconi}, {Hynes}, {Charles}, \& {Stella}}]{cdc+04}
{Campana}, S., {D'Avanzo}, P., {Casares}, J., {Covino}, S., {Israel}, G.,
  {Marconi}, G., {Hynes}, R., {Charles}, P., \& {Stella}, L. 2004, ApJ, 614,
  L49

\bibitem[{{Campana} {et~al.}(2002){Campana}, {Stella}, {Gastaldello},
  {Mereghetti}, {Colpi}, {Israel}, {Burderi}, {Di Salvo}, \& {Robba}}]{csg+02}
{Campana}, S., {Stella}, L., {Gastaldello}, F., {Mereghetti}, S., {Colpi}, M.,
  {Israel}, G.~L., {Burderi}, L., {Di Salvo}, T., \& {Robba}, R.~N. 2002,
  \apjl, 575, L15

\bibitem[{Chakrabarty \& Morgan(1998)}]{cm98}
Chakrabarty, D. \& Morgan, E.~H. 1998, Nature, 394, 346

\bibitem[{D'Amico {et~al.}(2001)D'Amico, Lyne, Manchester, Possenti, \&
  Camilo}]{dlm+01}
D'Amico, N., Lyne, A.~G., Manchester, R.~N., Possenti, A., \& Camilo, F. 2001,
  ApJ, 548, L171

\bibitem[{{D'Amico} {et~al.}(2001){D'Amico}, {Possenti}, {Manchester},
  {Sarkissian}, {Lyne}, \& {Camilo}}]{dpm+01}
{D'Amico}, N., {Possenti}, A., {Manchester}, R.~N., {Sarkissian}, J., {Lyne},
  A.~G., \& {Camilo}, F. 2001, ApJ, 561, L89

\bibitem[{{de Jager} {et~al.}(1989){de Jager}, {Raubenheimer}, \&
  {Swanepoel}}]{jrs89}
{de Jager}, O.~C., {Raubenheimer}, B.~C., \& {Swanepoel}, J.~W.~H. 1989, \aap,
  221, 180

\bibitem[{{Deloye}(2008)}]{delo08}
{Deloye}, C.~J. 2008, in American Institute of Physics Conference Series, Vol.
  983, 40 Years of Pulsars: Millisecond Pulsars, Magnetars and More, ed.
  C.~{Bassa}, Z.~{Wang}, A.~{Cumming}, \& V.~M. {Kaspi}, 501--509

\bibitem[{{Ferraro} {et~al.}(2001){Ferraro}, {Possenti}, {D'Amico}, \&
  {Sabbi}}]{fpds01}
{Ferraro}, F.~R., {Possenti}, A., {D'Amico}, N., \& {Sabbi}, E. 2001, \apjl,
  561, L93

\bibitem[{{Freire} {et~al.}(2003){Freire}, {Camilo}, {Kramer}, {Lorimer},
  {Lyne}, {Manchester}, \& {D'Amico}}]{fck+03}
{Freire}, P.~C., {Camilo}, F., {Kramer}, M., {Lorimer}, D.~R., {Lyne}, A.~G.,
  {Manchester}, R.~N., \& {D'Amico}, N. 2003, \mnras, 340, 1359

\bibitem[{{Fruchter} {et~al.}(1988){Fruchter}, {Stinebring}, \&
  {Taylor}}]{fst88}
{Fruchter}, A.~S., {Stinebring}, D.~R., \& {Taylor}, J.~H. 1988, Nature, 333,
  237

\bibitem[{{Gaensler} \& {Slane}(2006)}]{gs06}
{Gaensler}, B.~M. \& {Slane}, P.~O. 2006, \araa, 44, 17

\bibitem[{{Galloway} \& {Cumming}(2006)}]{gc06}
{Galloway}, D.~K. \& {Cumming}, A. 2006, \apj, 652, 559

\bibitem[{{Guainazzi} {et~al.}(2010)}]{g+10}
{Guainazzi}, M. {et~al.} 2010, \emph{XMM-Newton} Science Operations Centre,
  XMM-SOC-CAL-TN-0018

\bibitem[{{Heinke} {et~al.}(2009){Heinke}, {Jonker}, {Wijnands}, {Deloye}, \&
  {Taam}}]{hjw+09}
{Heinke}, C.~O., {Jonker}, P.~G., {Wijnands}, R., {Deloye}, C.~J., \& {Taam},
  R.~E. 2009, ApJ, 691, 1035

\bibitem[{{Homer} {et~al.}(2006){Homer}, {Szkody}, {Chen}, {Henden}, {Schmidt},
  {Anderson}, {Silvestri}, \& {Brinkmann}}]{hsc+06}
{Homer}, L., {Szkody}, P., {Chen}, B., {Henden}, A., {Schmidt}, G., {Anderson},
  S.~F., {Silvestri}, N.~M., \& {Brinkmann}, J. 2006, AJ, 131, 562

\bibitem[{{Iacolina} {et~al.}(2009){Iacolina}, {Burgay}, {Burderi}, {Possenti},
  \& {di Salvo}}]{ibb+09}
{Iacolina}, M.~N., {Burgay}, M., {Burderi}, L., {Possenti}, A., \& {di Salvo},
  T. 2009, \aap, 497, 445

\bibitem[{{Kalberla} {et~al.}(2005){Kalberla}, {Burton}, {Hartmann}, {Arnal},
  {Bajaja}, {Morras}, \& {P{\"o}ppel}}]{kbh+05}
{Kalberla}, P.~M.~W., {Burton}, W.~B., {Hartmann}, D., {Arnal}, E.~M.,
  {Bajaja}, E., {Morras}, R., \& {P{\"o}ppel}, W.~G.~L. 2005, \aap, 440, 775

\bibitem[{{Kargaltsev} \& {Pavlov}(2008)}]{kp08}
{Kargaltsev}, O. \& {Pavlov}, G.~G. 2008, in American Institute of Physics
  Conference Series, Vol. 983, 40 Years of Pulsars: Millisecond Pulsars,
  Magnetars and More, ed. {C.~Bassa, Z.~Wang, A.~Cumming, \& V.~M.~Kaspi},
  171--185

\bibitem[{{Kaspi} {et~al.}(2006){Kaspi}, {Roberts}, \& {Harding}}]{krh06}
{Kaspi}, V.~M., {Roberts}, M.~S.~E., \& {Harding}, A.~K. 2006, {Isolated
  neutron stars}, ed. {Lewin, W.~H.~G.~\& van der Klis, M.} (Cambridge
  University Press), 279--339

\bibitem[{{Paltani}(2004)}]{palt04}
{Paltani}, S. 2004, \aap, 420, 789

\bibitem[{{Protassov} {et~al.}(2002){Protassov}, {van Dyk}, {Connors},
  {Kashyap}, \& {Siemiginowska}}]{pdc+02}
{Protassov}, R., {van Dyk}, D.~A., {Connors}, A., {Kashyap}, V.~L., \&
  {Siemiginowska}, A. 2002, \apj, 571, 545

\bibitem[{Stappers {et~al.}(1996)Stappers, Bailes, Lyne, Manchester, D'Amico,
  Tauris, Lorimer, Johnston, \& Sandhu}]{sbl+96}
Stappers, B.~W., Bailes, M., Lyne, A.~G., Manchester, R.~N., D'Amico, N.,
  Tauris, T.~M., Lorimer, D.~R., Johnston, S., \& Sandhu, J.~S. 1996, ApJ, 465,
  L119

\bibitem[{{Stella} {et~al.}(1994){Stella}, {Campana}, {Colpi}, {Mereghetti}, \&
  {Tavani}}]{scc+94}
{Stella}, L., {Campana}, S., {Colpi}, M., {Mereghetti}, S., \& {Tavani}, M.
  1994, \apjl, 423, L47

\bibitem[{{Stewart}(2009)}]{stew09}
{Stewart}, I.~M. 2009, \aap, 495, 989

\bibitem[{{Str{\"u}der} {et~al.}(2001){Str{\"u}der}, {Briel}, {Dennerl},
  {Hartmann}, {Kendziorra}, {Meidinger}, {Pfeffermann}, {Reppin}, {Aschenbach},
  {Bornemann}, {Br{\"a}uninger}, {Burkert}, {Elender}, {Freyberg}, {Haberl},
  {Hartner}, {Heuschmann}, {Hippmann}, {Kastelic}, {Kemmer}, {Kettenring},
  {Kink}, {Krause}, {M{\"u}ller}, {Oppitz}, {Pietsch}, {Popp}, {Predehl},
  {Read}, {Stephan}, {St{\"o}tter}, {Tr{\"u}mper}, {Holl}, {Kemmer}, {Soltau},
  {St{\"o}tter}, {Weber}, {Weichert}, {von Zanthier}, {Carathanassis}, {Lutz},
  {Richter}, {Solc}, {B{\"o}ttcher}, {Kuster}, {Staubert}, {Abbey}, {Holland},
  {Turner}, {Balasini}, {Bignami}, {La Palombara}, {Villa}, {Buttler},
  {Gianini}, {Lain{\'e}}, {Lumb}, \& {Dhez}}]{sbd+01}
{Str{\"u}der}, L., {Briel}, U., {Dennerl}, K., {Hartmann}, R., {Kendziorra},
  E., {Meidinger}, N., {Pfeffermann}, E., {Reppin}, C., {Aschenbach}, B.,
  {Bornemann}, W., {Br{\"a}uninger}, H., {Burkert}, W., {Elender}, M.,
  {Freyberg}, M., {Haberl}, F., {Hartner}, G., {Heuschmann}, F., {Hippmann},
  H., {Kastelic}, E., {Kemmer}, S., {Kettenring}, G., {Kink}, W., {Krause}, N.,
  {M{\"u}ller}, S., {Oppitz}, A., {Pietsch}, W., {Popp}, M., {Predehl}, P.,
  {Read}, A., {Stephan}, K.~H., {St{\"o}tter}, D., {Tr{\"u}mper}, J., {Holl},
  P., {Kemmer}, J., {Soltau}, H., {St{\"o}tter}, R., {Weber}, U., {Weichert},
  U., {von Zanthier}, C., {Carathanassis}, D., {Lutz}, G., {Richter}, R.~H.,
  {Solc}, P., {B{\"o}ttcher}, H., {Kuster}, M., {Staubert}, R., {Abbey}, A.,
  {Holland}, A., {Turner}, M., {Balasini}, M., {Bignami}, G.~F., {La
  Palombara}, N., {Villa}, G., {Buttler}, W., {Gianini}, F., {Lain{\'e}}, R.,
  {Lumb}, D., \& {Dhez}, P. 2001, \aap, 365, L18

\bibitem[{{Thorstensen} \& {Armstrong}(2005)}]{ta05}
{Thorstensen}, J.~R. \& {Armstrong}, E. 2005, AJ, 130, 759

\bibitem[{{Turner} {et~al.}(2001){Turner}, {Abbey}, {Arnaud}, {Balasini},
  {Barbera}, {Belsole}, {Bennie}, {Bernard}, {Bignami}, {Boer}, {Briel},
  {Butler}, {Cara}, {Chabaud}, {Cole}, {Collura}, {Conte}, {Cros}, {Denby},
  {Dhez}, {Di Coco}, {Dowson}, {Ferrando}, {Ghizzardi}, {Gianotti}, {Goodall},
  {Gretton}, {Griffiths}, {Hainaut}, {Hochedez}, {Holland}, {Jourdain},
  {Kendziorra}, {Lagostina}, {Laine}, {La Palombara}, {Lortholary}, {Lumb},
  {Marty}, {Molendi}, {Pigot}, {Poindron}, {Pounds}, {Reeves}, {Reppin},
  {Rothenflug}, {Salvetat}, {Sauvageot}, {Schmitt}, {Sembay}, {Short},
  {Spragg}, {Stephen}, {Str{\"u}der}, {Tiengo}, {Trifoglio}, {Tr{\"u}mper},
  {Vercellone}, {Vigroux}, {Villa}, {Ward}, {Whitehead}, \& {Zonca}}]{taa+01}
{Turner}, M.~J.~L., {Abbey}, A., {Arnaud}, M., {Balasini}, M., {Barbera}, M.,
  {Belsole}, E., {Bennie}, P.~J., {Bernard}, J.~P., {Bignami}, G.~F., {Boer},
  M., {Briel}, U., {Butler}, I., {Cara}, C., {Chabaud}, C., {Cole}, R.,
  {Collura}, A., {Conte}, M., {Cros}, A., {Denby}, M., {Dhez}, P., {Di Coco},
  G., {Dowson}, J., {Ferrando}, P., {Ghizzardi}, S., {Gianotti}, F., {Goodall},
  C.~V., {Gretton}, L., {Griffiths}, R.~G., {Hainaut}, O., {Hochedez}, J.~F.,
  {Holland}, A.~D., {Jourdain}, E., {Kendziorra}, E., {Lagostina}, A., {Laine},
  R., {La Palombara}, N., {Lortholary}, M., {Lumb}, D., {Marty}, P., {Molendi},
  S., {Pigot}, C., {Poindron}, E., {Pounds}, K.~A., {Reeves}, J.~N., {Reppin},
  C., {Rothenflug}, R., {Salvetat}, P., {Sauvageot}, J.~L., {Schmitt}, D.,
  {Sembay}, S., {Short}, A.~D.~T., {Spragg}, J., {Stephen}, J., {Str{\"u}der},
  L., {Tiengo}, A., {Trifoglio}, M., {Tr{\"u}mper}, J., {Vercellone}, S.,
  {Vigroux}, L., {Villa}, G., {Ward}, M.~J., {Whitehead}, S., \& {Zonca}, E.
  2001, \aap, 365, L27

\bibitem[{{Wang} {et~al.}(2009){Wang}, {Archibald}, {Thorstensen}, {Kaspi},
  {Lorimer}, {Stairs}, \& {Ransom}}]{wat+09}
{Wang}, Z., {Archibald}, A.~M., {Thorstensen}, J.~R., {Kaspi}, V.~M.,
  {Lorimer}, D.~R., {Stairs}, I., \& {Ransom}, S.~M. 2009, \apj, 703, 2017

\bibitem[{{Wijnands} \& {van der Klis}(1998)}]{wk98}
{Wijnands}, R. \& {van der Klis}, M. 1998, Nature, 394, 344

\bibitem[{{Zacharias} {et~al.}(2004){Zacharias}, {Monet}, {Levine}, {Urban},
  {Gaume}, \& {Wycoff}}]{zml+04}
{Zacharias}, N., {Monet}, D.~G., {Levine}, S.~E., {Urban}, S.~E., {Gaume}, R.,
  \& {Wycoff}, G.~L. 2004, in Bulletin of the American Astronomical Society,
  Vol.~36, Bulletin of the American Astronomical Society, 1418

\bibitem[{{Zavlin}(2007)}]{zavl07}
{Zavlin}, V.~E. 2007, \apss, 308, 297

\bibitem[{{Zavlin} {et~al.}(1996){Zavlin}, {Pavlov}, \& {Shibanov}}]{zps96}
{Zavlin}, V.~E., {Pavlov}, G.~G., \& {Shibanov}, Y.~A. 1996, \aap, 315, 141

\end{thebibliography}

\end{document}